\documentclass[letterpaper, 10 pt, conference]{ieeeconf}

\IEEEoverridecommandlockouts                              

\usepackage{cite}

  \usepackage[pdftex]{graphicx}

\usepackage{amsmath}
\usepackage{amssymb}

\title{\LARGE \bf 
Synthesis of Different Autonomous Vehicles Test Approaches}

\author{Zhiyuan Huang$^{1}$, 
Mansur Arief$^{2}$, Henry Lam$^{3}$, and Ding Zhao$^{2}$
\thanks{We gratefully acknowledge support from the National Science Foundation under grants CMMI-1542020 and CAREER CMMI-1653339. }
\thanks{$^{1}$Zhiyuan Huang is with the Department of Industrial and Operations Engineering at
University of Michigan, 
1205 Beal Ave, MI, USA
        {\tt\small zhyhuang@umich.edu}  }%
\thanks{$^{2}$ Mansur Arief and Ding Zhao is with Carnegie Mellon University, USA
        {\tt\small marief@andrew.cmu.edu
, dingzhao@cmu.edu} }%
\thanks{$^{3}$Henry Lam are with the Department of Industrial Engineering and Operations Research, Columbia University,
        500 W. 120th Street, NY, USA
        {\tt\small khlam@umich.edu}}%
}

\begin{document}

\maketitle
\thispagestyle{empty}
\pagestyle{empty}

\begin{abstract}
Currently, the most prevalent way to evaluate an autonomous vehicle is to directly test it on the public road. However, because of recent accidents caused by autonomous vehicles, it becomes controversial about whether on-road tests should be the best approach. Alternatively, people use test tracks or simulation to assess the safety of autonomous vehicles. These approaches are time-efficient and less costly, however, their credibility varies. In this paper, we propose to use a co-Kriging model to synthesize the results from different evaluation approaches, which allows us to fully utilize the information and provides an accurate, affordable, and safe way to assess a design of an autonomous vehicle.
\end{abstract}


\section{Introduction}

Recently, a pedestrian was killed by a self-driving car in a crash in Arizona \cite{Wakabayashi2018}. This accident has brought up a hot debate on whether it is safe to test autonomous vehicles (AVs) on public roads. People argue that it is not responsible to test self-driving cars in public areas, because of the safety concern. On the other hand, AVs are designed to operate on public roads. To claim an AV has a high safety level without on-road tests is not fully convincing.

Besides the safety issue, on-road tests are also time-consuming and expensive to implement. People have already considered alternatives to on-road tests. For instance, Waymo developed computer simulation platform for self-driving cars training and testing \cite{madrigal_2017}. Studies also consider how to smartly utilize test tracks, e.g. combining with Augmented Reality (AR) technique, which generates virtual cars and pedestrians on a test track.   

These test approaches provides plenty of choices, however, it is hard to claim either one is the ``best'' option. The pros and cons of these approaches make it unclear that what test should be implemented if only one of them can be chosen. The on-road tests are expensive and risky, but it is the most credible. The AR on-track test is less risky, but it might overlook factors in naturalistic driving environment and is still relatively time-consuming and expensive. Computer simulation is cheap and quick, however, it is less credible than physical tests. On the other hand, test results from different resources cannot be naively combined because of the different credibility.

The nature of autonomous and intelligent systems also brings difficulties in evaluating AVs from historical performance. AVs are based on algorithms that might be updated after a while, which makes it illegitimate to directly utilize historical test results and data that is collected before the update. Consider that these data contains some information about the current version of algorithms and is relatively plenty in general, the evaluation procedure will be more efficient if we have a way to link them with the current algorithm.

In this paper, we are targeting this problem in AV testing: how to synthesize ``independent'' test results from different test approaches and historical data from outdated models into the safety evaluation of an AV. We discuss an AV evaluation approach that is capable of synthesizing and integrating test results from different resources and historical data. This approach is based on co-Kriging models, in which we consider each type of tests as models with an assigned fidelity level (we take historical data as a type of test). The model provides a response surface for the test performance function of interest. The model allows us to analyze the performance of a test AV using a combination of different types of test results (and historical test data). Furthermore, the proposed model can potentially be used to design new experiments, i.e, determine what test to implement based on current information for improving the model with regards to evaluation accuracy. 

To link the proposed model with AV testing, we follow the test framework in the accelerated evaluation method, which is first proposed in \cite{Zhao2016AcceleratedTechniques}. We decompose naturalistic driving environment into different test scenarios and use statistical models to represent the environment in the scenarios. For each test scenario, we use the probability of safety-critical events to evaluate the safety level of an AV. \cite{huang2017towards} discussed an approach that uses response surface model in the accelerated evaluation context, and used a single type test results.  This paper extends the use of response surface model in \cite{huang2017towards}. Besides results of on-road tests, the proposed approach also obtains information from other test resources, which improves the accuracy of the response surface. 

The framework we proposed is related to multi-fidelity models (for review, see \cite{fernandez2016review}). The co-Kriging model we adopt is originally considered in \cite{forrester2007multi}, which was proposed in an optimization context. This model utilizes Kriging model or Gaussian process in constructing the response surface. We also discuss the extension of the design of experiments scheme in \cite{huang2017towards} to multi-source tests. The scheme can help us smartly select design points and therefore avoid unnecessary experiments in the model constructing procedure. We will further illustrate the difference between these two approaches using several numerical examples. We use the proposed methods to study the lane change scenario, which has been studied in \cite{Zhao2016AcceleratedTechniques,Huang2016AcceleratedModels,huang2017versatile}. 


This paper is structured as follows: Section \ref{sec:basics} introduce the basics of Kriging, co-Kriging and the multi-fidelity models. Section \ref{sec:AE} discusses the properties of the proposed multi-fidelity model, the design of experiment scheme and the application in AV testing.  Section \ref{sec:numerical} shows numerical experiments using the proposed method. 

\section{Kriging-based Surrogate Models}
\label{sec:basics}

In this section, we introduce the Kriging-based surrogate models that we propose to use for AV testing. We first review the basics of Kriging model. Then we introduce the idea of co-Kriging and show how to turn this idea into a multi-fidelity model. The model is extended from \cite{forrester2007multi}.

\subsection{Basics of Kriging Model}

\label{sec:kriging}

Kriging, a model named after the developer Krige, was originally used in geostatistics \cite{krige1966two}. The model has been extensively used in engineering fields since it is studied and introduced under the design of experiments context in \cite{sacks1989design}. For a wider scope of Kriging application, one can refer to \cite{Kleijnen2009KrigingReview}.

Suppose we want to study a performance function $g(x)$ on the design space $x\in \mathcal X \subseteq \mathbb{R}^d$. The performance function is only available through experiments (or observations). The Kriging model allows us to construct a response surface based on experiment results, $(x^i,g(x^i))$ for $i=1,...,n$, where $n$ is the number of experiments we have collected.

Here we consider a Bayesian view of Kriging model. The key idea of Kriging is to consider the response surface of $g(x)$ as a posterior of a Gaussian random field (or Gaussian process) \cite{Rasmussen2004GaussianLearning,Staum2009BetterKriging}. A Gaussian random field $y(x)$ for $x\in \mathbb{R}^d$ is specified by a mean function, $\mu(x)$, and a covariance function, $\sigma^2(x,x')$. We denote such a Gaussian random field as \begin{equation}
	y \sim GRF(\mu,\sigma^2).
\end{equation}
For any $x\in \mathcal X$, $y(x)$ is Gaussian random variable with mean $\mu(x)$ and variance $\sigma^2(x,x)$. For $x,x' \in \mathbb{R}^d $, the covariance between $y(x)$ and $y(x')$ is $\sigma^2(x,x')$. We assume the following structure for the mean and covariance function $\mu(x)=b(x) \beta$ and $\sigma^2(x,x')=\tau^2 r(x,x';\theta)$, where $\beta$, $\tau^2$ and $\theta$ are tunable parameters. Note that the covariance function indicates that the variance $\tau^2$ is stationary over $x$ .

We consider the above functions as the prior mean and covariance of the Gaussian random field $y(x)$. Let $X$ denotes the experiments at $\{x^1,...,x^n\}$ and $Y$ denotes the corresponding observations$\{y^1,...,y^n\}$. We use $X$ to construct a matrix $\Sigma$, where $\Sigma_{ij}=\sigma^2(x^i,x^j)$. And let $R=\Sigma/\tau^2$. Note that $R_{ij}=r(x^i,x^j;\theta)$. Given observations $(X,Y)$, for any $x\in \mathbb{R}^d$ we have the posterior mean and covariance function as \begin{equation} 
	E(y(x)|X,Y)= \mu(x) +r(x)'R^{-1}(Y-\mu(x))
	\label{eq:kriging_E}
\end{equation}
and
\begin{equation} 
	Var(y(x)|X,Y)=\tau^2 (1- r(x)'R^{-1}r(x)),
	\label{eq:kriging_var}
\end{equation}
where $r(x)$ returns a vector with $r(x,x^i)$ as the $i$th element.

This posterior Gaussian random field $y(x)|X,Y$ is the Kriging model for $g(x)$. In this paper, we use $\mu(x)=\beta$, $\beta \in \mathbb{R}$. We denote  $\mu(x|X,Y)=E(y(x)|X,Y)$ and  $\sigma^2(x|X,Y)=Var(y(x)|X,Y)$ for simplification.

For choosing the tunable parameters, $\beta$, $\tau^2$ and $\theta$, in the prior, one can use maximum likelihood estimation (MLE) using data, i.e: \begin{equation}
\hat{\beta}= \frac{\sum_{i=1}^{n}y_i}{n},
\end{equation}
 and we maximize the log likelihood function \begin{equation}
	l(\tau^2,\theta)=-\frac{1}{2} \left( n \log(2\pi) +\log(|\Sigma|) + (Y-\beta)' \Sigma^{-1}  (Y-\beta)   \right)
 \end{equation}
for $\hat{\tau}^2$ and $\hat{\theta}$. For more details about the MLE estimator, see, e.g., \cite{ankenman2010stochastic}.

\subsection{Co-Kriging and Multi-fidelity Model}\label{sec:cokriging}

\subsubsection{Co-Kriging}
Here we discuss the co-Kriging model that is studied in \cite{forrester2007multi} and show how to extend it to fit for AV testing. The idea of co-Kriging is to use the summation of two Kriging models as the response surface. Because of a nice property for Gaussian random variables (the summation of two Gaussian random variables is still Gaussian), the co-Kriging model is still a Gaussian random field. 

Now we consider that the performance function $g(x)$ is a summation of two factor functions $g_1(x)$ and $g_2(x)$, i.e. $g(x)=g_1(x)+g_2(x)$. And the factor functions are only available through experiments. We use $(X_1,Y_1)$ to denote the data set for the factor function $g_1(x)$, where $X_1$ contains experiments $\{x^1,...,x^{n_1}\}$ and $Y_1$ contains the corresponding observations $\{g_1(x^1),...,g_1(x^{n_1})\}$. Similarly, $(X_2,Y_2)$ denotes the data set for factor function $g_2(x)$ with $n_2$ observations. We use $(X,Y)$ to denote the whole data set (including $(X_1,Y_1)$ and $(X_2,Y_2)$). 

We construct Kriging models that is described in Section \ref{sec:kriging} for both factor functions $g_1(x)$ and $g_2(x)$ independently. (Here we assume the two factors are independent, which means that the value of $g_1(x)$ does not contain any information for $g_2(x)$.) We denote the Kriging model for $g_1(x)$ as $y_1(x)$ and the Kriging model for $g_2(x)$ as $y_2(x)$. The co-Kriging model for the performance function $g(x)$ is given by $y(x)=y_1(x)+y_2(x)$. 

As we mentioned, the co-Kriging model $y(x)$ is still a Gaussian random field. For any $x\in \mathcal X$, $y(x)$ is the summation of two Gaussian random variable $y_1(x)$ and $y_2(x)$. Therefore, $y(x)$ has mean $\mu(x|(X,Y))=\mu_1(x|(X_1,Y_1))+\mu_2(x|(X_2,Y_2))$ and variance $\sigma^2(x,x|(X,Y))=\sigma^2_1(x,x|(X_1,Y_1))+\sigma^2_2(x,x|(X_2,Y_2))$ (because $y_1$ and $y_2$ are independent), where $\mu_1,\mu_2$ denotes the mean function for $y_1$ and $y_2$ respectively , and $\sigma^2_1,\sigma^2_2$ denotes the covariance function for $y_1$ and $y_2$ respectively.

\subsubsection{Multi-fidelity Model}
To extend the general co-Kriging model shown above as a multi-fidelity model, we consider the following settings. We consider that we have $T$ models with different fidelity for the performance function $g(x)$, which are denoted as $h_t(x),t=1,...,T$, where $t$ is the fidelity level. We assume that a larger $t$ indicates a better fidelity. In this case, $h_1(x)$ is the model with lowest fidelity and $h_T(x)$ is the model with the highest fidelity. Usually, we consider $h_T(x)=g(x)$, which means that the original performance function is the highest fidelity model. 

For each fidelity $t$, we have a data set $(X_t,Y_t)$, where $X_t$ contains $n_t$ experiments on $\{x^1,...,x^{n_t}\}$ and $Y_t$ contains the corresponding observations $\{ h_t(x^1),..., h_t(x^{n_t})\}$. Here we adopt the assumption in \cite{forrester2007multi}, that is $X_T \subseteq X_{T-1} \subseteq ...\subseteq X_1$. This assumption is not necessary for the construction of the multi-fidelity model, however, it allows the multi-fidelity model to maintain a nice property of single Kriging model (further discussed in Section \ref{sec:remarks}). We use $(X,Y)$ to denote all data sets for simplicity.

The procedure of constructing the multi-fidelity model is as follows. We first construct a Kriging model for the lowest fidelity model $h_1(x)$ using data set $(X_1,Y_1)$ and we denote the model as $y_1(x)$. We have $y_1(x)$ as a Gaussian random field with mean \begin{equation}
	\mu_1(x|X_1,Y_1)= \mu_1(x) +r_1(x)'R_1^{-1}(Y_1-\mu_1(x))
\end{equation}
and variance
\begin{equation}
	\sigma^2_1(x|X_1,Y_1)=\tau^2_1 (1- r_1(x)'R_1^{-1}r_1(x)).
\end{equation}
Note that the notation for parameters and functions follows \eqref{eq:kriging_E} and \eqref{eq:kriging_var} in Section \ref{sec:kriging}. 

Then we construct response surface for other fidelities layer by layer. Starting from the fidelity $t=2$, we consider to build a Kriging model for the difference between two adjacent fidelities, i.e. $h_t(x)-h_{t-1}(x)$. We create a data set $(X_t,D_t)$ using $(X_{t-1},Y_{t-1})$ and $(X_t,Y_t)$, such that $D_t=\{h_t(x^1)-h_{t-1}(x^1),...,h_t(x^{n_t})-h_{t-1}(x^{n_t})\}$. (For any $x\in X_t$, we have $h_{t-1}(x)$ because $X_t \subseteq X_{t-1}$.) Now we use the created data set $(X_t,D_t)$ to construct a Kriging model, that we denote as $d_t(x)$. Note that $d_t(x)$ is a Gaussian random field with mean (for simplicity we still use $(X_t,Y_t)$ to denote the data set we use) \begin{equation}
	\mu_t(x|X_t,Y_t)= \mu_t(x) +r_t(x)'R_t^{-1}(Y_t-\mu_t(x))
\end{equation}
and variance
\begin{equation}\label{eq:layer_var}
	\sigma^2_t(x|X_t,Y_t)=\tau^2_t (1- r_t(x)'R_t^{-1}r_t(x)).
\end{equation}
Then we have the response surface model $y_t(x)$ for $h_t(x)$, which is given by $y_t(x)=y_{t-1}(x) + d_t(x)$. For convenience, we have define $d_1(x)=y_1(x)$. Now for the model with fidelity $t$, we have \begin{equation}
y_t(x)= \sum_{i=1}^{t} d_i(x).
\end{equation}
Note that each $d_i(x)$ is a Gaussian random field, and therefore $y_t(x)$ is still a Gaussian random field. 

$y_T(x)$ is our multi-fidelity model for the performance function $g(x)$, which is a Gaussian random field with mean function \begin{equation}
	\mu_T(x|X,Y)= \sum_{i=1}^{T} \mu_i(x|X_i,Y_i)
\end{equation}
and variance function
\begin{equation}\label{eq:final_var}
	\sigma^2_T(x|X,Y)=\sum_{i=1}^{T} \sigma^2_i(x|X_i,Y_i).
\end{equation}

\subsection{Remarks}\label{sec:remarks}

Compared to Kriging model that only uses observations of the performance function $g(x)$ (or the highest fidelity model), the proposed multi-fidelity model integrates information from models with lower fidelities, while it maintains an important property of the Kriging model. In Kriging, the prediction on $x$ is exact if you have already observed $x$. With the assumption on data set structure, the proposed multi-fidelity model maintains this good property. This means that the prediction accuracy of the proposed multi-fidelity model will increase in a similar way as the Kriging model and observations of $g(x)$ increases.

Since the proposed multi-fidelity model provides a linkage between the performance function and the lower fidelity models, we are able to study how much information an experiment in lower fidelity can bring. This linkage allows us to design experiments and choose the fidelity level that is economic with regard to the information it brings in. 

As a side product, the proposed multi-fidelity model provides a response surface model for each fidelity model. This side product allows us to develop experiment design scheme that uses lower fidelity information. We will further discuss this in Section \ref{sec:AE}.

The response surface models for different fidelity levels have a property as follows. For any fidelity parameters $t > t'$ and at $x$, we always have $Var(y_t(x)) \geq Var(y_{t'} (x))$ because $Var(y_t(x))  = \sum_{i=1}^{t} \sigma^2_i(x|X_i,Y_i)$ and $\sigma^2_i(x|X_i,Y_i) \geq 0$. This property is intuitively reasonable, since we have less information about a higher fidelity model.

\section{Synthesizing Tests in Accelerated Evaluation}\label{sec:AE}

In this section, we discuss applying the proposed model to AV testing. More specifically, we consider applying the proposed model in the context of test scenario based AV evaluation that has been studied by \cite{Zhao2016AcceleratedTechniques,huang2017towards}. We first review the problem setting in test scenario based AV evaluation. We then show how this model is applied to synthesize data from different test sources. 

\subsection{Problem Setting in AV Evaluation} \label{sec:setting}

Accelerated evaluation \cite{Zhao2016AcceleratedTechniques} is an approach to efficiently evaluate the safety level of an AV. This approach evaluates AV based on the test AV's performance in different traffic scenarios. The traffic scenarios are decomposed from naturalistic driving and are considered to be safety-critical since a very high percentage of crashes occurred in these scenarios \cite{najm2013depiction}. Examples of there scenarios are discussed in \cite{Huang2016AcceleratedModels,ZhaoLeftTurn,ZhaoPedes}.

For each of these traffic scenarios, the uncertainty in the driving environment is modeled as stochastic (follows some statistical model). The probability of safety-critical events (e.g. crash) is used as the criterion for determining the safety level. Therefore, the task of this approach is to estimate the probability of safety critical events in each test scenarios.

The problem is mathematically defined as follows. Let us use $x \in \mathbb R$ to denote the variable that represents the uncertainty in the driving environment and $x$ is modeled as a distribution $f(x)$. We use $g(x)$ to represent a performance function that the safety-critical events depend on and use $\gamma$ to denote the threshold for $g(x)$ to trigger the safety-critical events (this means that $g(x) \geq \gamma$ indicates that a safety-critical event occurs at $x$). 

\begin{figure}[t]
      \centering
      \includegraphics[width=\linewidth]{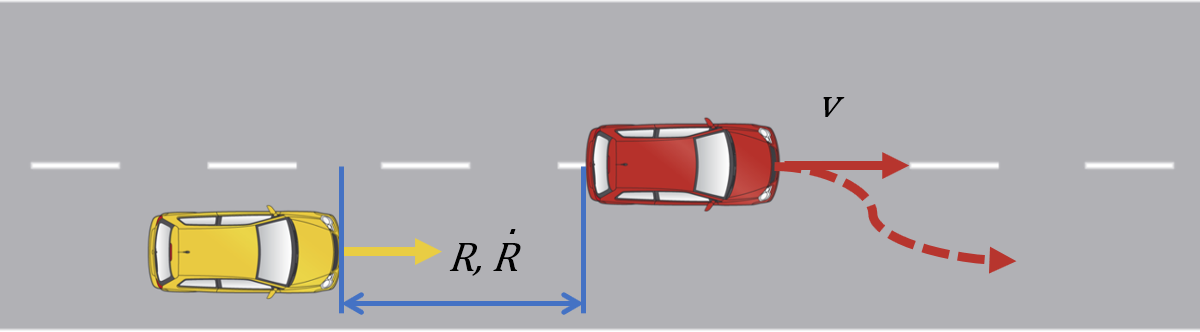}
      \caption{The lane change test scenario.}
      \label{fig:laneChange}
\end{figure}

For instance, if we want to estimate the probability of crash in the lane change scenario (refer to Fig. \ref{fig:laneChange}). In this test scenario, a frontal human-driving vehicle cuts into the lane of a test AV. The environment uncertainty, $x$, consists of the following variables: the velocity of the frontal car, $v$, the relative speed between the frontal car and the test AV, $\dot{R}$, and the range between these two cars, $R$. The uncertainty of these variables is modeled as probability distribution $f(x)$. We define the performance function $g(x)$ as the minimum range between the test AV and the frontal vehicle. We estimate $P( \{x: g(x) <=0 \})$, which represents the probability of crash in this scenario, to evaluate the safety level of an AV under this test scenario.

\subsection{Synthesizing Tests}\label{sec:use_of_MFM}

When we want to evaluate an AV in a certain test scenario, we have several resources to select. Here we consider these different test resources as models with different fidelities. We rank the fidelity level of the test resources in an arbitrary way (the rank is unnecessary to be ``correct'' for the model to work, but will affect the accuracy). For example, we consider the on-road test as the highest fidelity model, since this is the ``true'' test in the evaluation. Then we consider an AR test has lower fidelity, because the AR test maintains the check on the physical part of the test AV. A pure computer simulation of the vehicle algorithm is considered to have lower fidelity than AR test, since physical parts are not considered in this case. Lastly, we consider the historical data or test results for similar designed AVs as the lowest fidelity model. 

\begin{figure}[t]
      \centering
      \includegraphics[width=\linewidth]{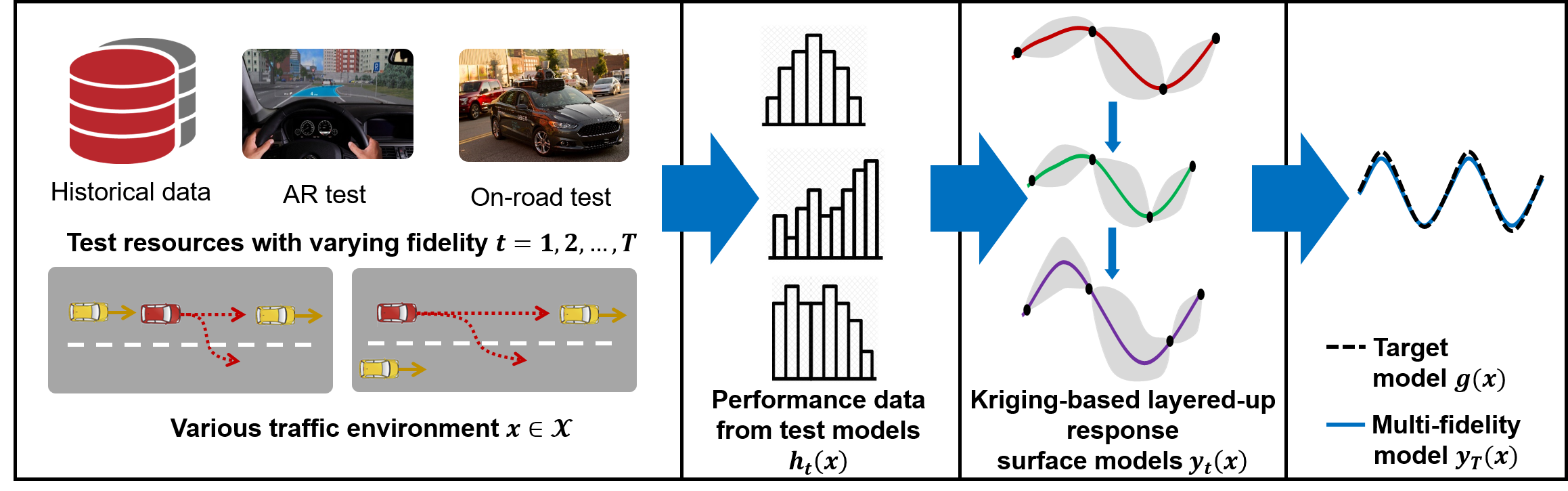}
      \caption{The proposed multi-fidelity model on AV safety evaluation.}
      \label{fig:mfm}
\end{figure}

After we rank different test resource with fidelity levels, given data set collected from these tests $(X_t,Y_t)$ for $t=1,...,T$, we are able to use the proposed multi-fidelity model to construct a response surface $y_T(x)$ for $g(x)$. The procedure follows the multi-fidelity model construction in Section \ref{sec:cokriging} and is illustrated in Fig. \ref{fig:mfm}. Following \cite{huang2017towards}, we use the response surface to estimate the probability of safety critical events. The estimation is given by \begin{equation}
\hat p=\hat P(g(x) \geq \gamma) = E_{x} \left[ P \left( y_T(x) \geq \gamma \right) \right],
\end{equation}
where the inner part $P \left( y_T(x) \geq \gamma \right)$ denotes the probability of $y_T(x) \geq \gamma$ given the value of $x$ and the outer expectation is over the distribution $f(x)$.

As we pointed out in Section \ref{sec:remarks}, besides the probability estimation, the multi-fidelity model can be used to provide a guideline for designing experiments. More specifically, here we want to collect new data $(X^{new},Y^{new})$ and use $(X^{new},Y^{new})\cup(X,Y)$ to construct a better model. We need to decide the fidelity levels $t$'s and the value of $x$ in $X^{new}$, so that we can do experiments on those $t$'s and $x$'s to collect the response $Y_{new}$. In \cite{huang2017towards,huang2017sequential}, an experiment design scheme for Kriging based on the information gain (IG) is discussed. Here, we define IG at point $x$ for model with fidelity $t$ as\begin{equation}
	IG(x,t)=E_{y \sim y_t(x)} \left[ (\hat p_{n} -\hat{p}_{n+1}(x,y))^2 | (X,Y)\right],
    \label{eq:IG_kriging}
\end{equation} 
where $\hat{p}$ denotes the probability estimation with $n$ samples in the samples set $(X,Y)$ and $\hat{p}_{n+1}(x,y)$ denotes the probability estimation with sample set $(X,Y)$ and an additional sample $(x,y)$. Here we use the response surface for the model with fidelity $t$ to compute the IG, where we take advantage of the ``side product'' of the multi-fidelity model. Consider that the cost for implement an experiment at $x$ for model with fidelity $t$ is $C(x,t)$, similar to \cite{stroh2017sequential}, we set our design selection criterion to be\begin{equation}
\label{eq:criterion}
(x,t)=\arg \max_{(x,t) } \frac{IG(x,t)}{C(x,t)}.
\end{equation}

\section{Numerical Experiments}\label{sec:numerical}

In this section, we consider two numerical experiments to show the advantage of the proposed model. We first consider a one-dimension problem to illustrate the proposed method. Then we apply the proposed model to an AV test scenario. 

\subsection{Illustration Example}\label{sec:toy}

To illustrate how the proposed model integrates experiment results from models with different fidelities, we set up an one-dimensional problem as follows. Suppose we are interested in the performance function $g(x)$ and we have two models $h_1(x)$ and $h_2(x)$ that are approximation of $g(x)$. For any design variable $x$, the response of $g(x), h_1(x), h_2(x)$ are unknown and need to be observed from an experiment at $x$. An experiment on the models $h_1(x), h_2(x)$ has a lower cost than an experiment on $g(x)$, while $h_1(x)$ has a lower cost but less accuracy (lower fidelity).    

\begin{figure}[t]
      \centering
      \includegraphics[width=\linewidth]{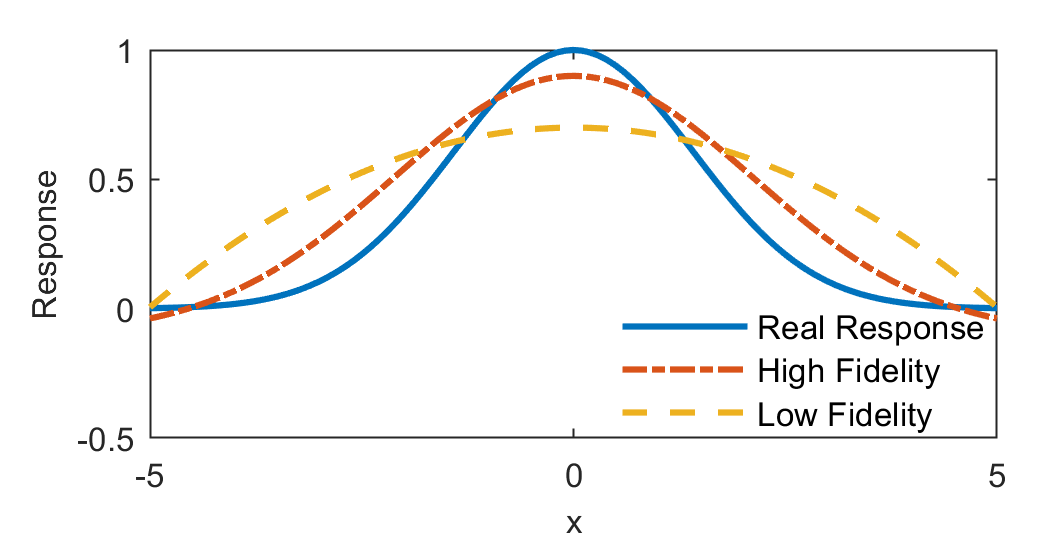}
      \caption{The response of the real performance function and two models with different fidelity.}
      \label{fig:exp1_g}
\end{figure}

In this example we assume that the real performance function of interest is\begin{equation}
g(x)= \exp{ \{ - \left(  \frac{x}{2}  \right)^2 \} }.
\end{equation} The model with higher fidelity is represented by \begin{equation}
h_2(x)= \exp{ \{ - \left(  \frac{x}{3}  \right)^2 \} }-0.1
\end{equation}
and the model with lower fidelity is given by\begin{equation}
h_1(x)= 0.7- \left( \frac{x}{6} \right) ^2. 
\end{equation} We consider the design space as $x\in [-5,5]$. Fig. \ref{fig:exp1_g} shows the response of these three functions at different $x$. We observe that the two models roughly capture the shape of the performance function, and the high fidelity is a better approximation to the performance function. Note that according to the notations we used in this paper, we have $g(x)=h_3(x)$.

With the above setting, let us consider that we have some experiment results from these models and we use these results to construct a response surface for the performance function. For the performance function $g(x)$, we have experiment results at $x=-5,-2,1,4$. For the higher fidelity model $h_2(x)$, we have experiment results at $x=-5,-3.5,-2,-0.5,1,2.5,4$. For the lower fidelity model $h_1(x)$, we have experiment results at $x=-5,-4.5,-4,...,4,4.5,5$. (We have more experiment results for lower fidelity models.)

\begin{figure}[t]
      \centering
      \includegraphics[width=\linewidth]{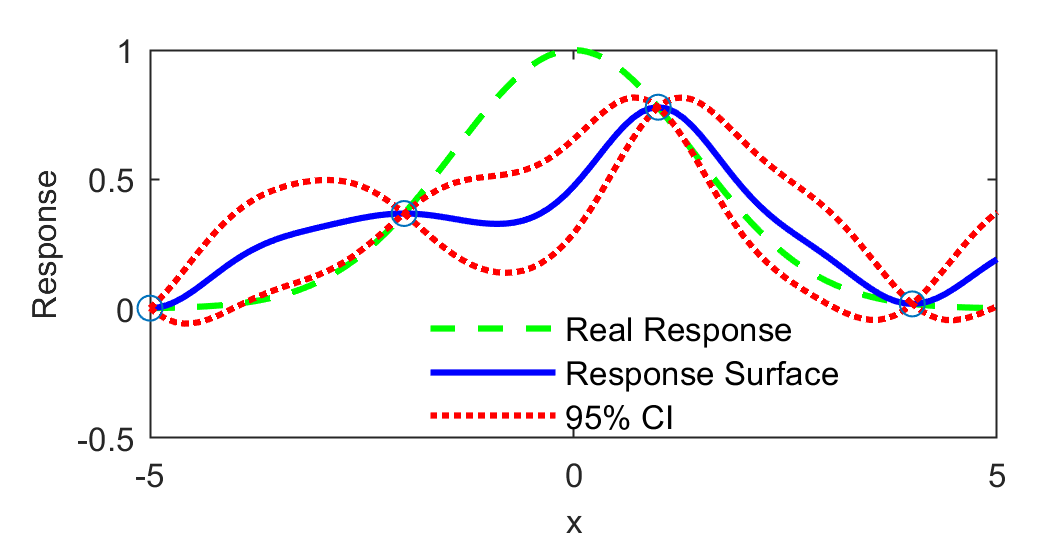}
      \caption{Response surface constructed from the experiments from the performance function $g(x)$. The blue circles represent the experiment results from $g(x)$.}
      \label{fig:exp1_1l}
\end{figure}

We first construct a Kriging model with the experiment results from the performance function. We use the Kriging model as the baseline of response surface models. Fig. \ref{fig:exp1_1l} shows the Kriging model we obtain. We observe that the mean of the response surface (blue solid line) is not close to the real function (green dash line) in most part of the region (e.g. $[-5,1]$ and $[4,5]$) and the 95\% confidence interval (red dot line) does not contain the real function in $[-1,3]$. Note that the Kriging model is built with only 4 data points (the blue circles in Fig. \ref{fig:exp1_1l}), the inaccuracy of the response surface is expected.

\begin{figure}[t]
      \centering
      \includegraphics[width=\linewidth]{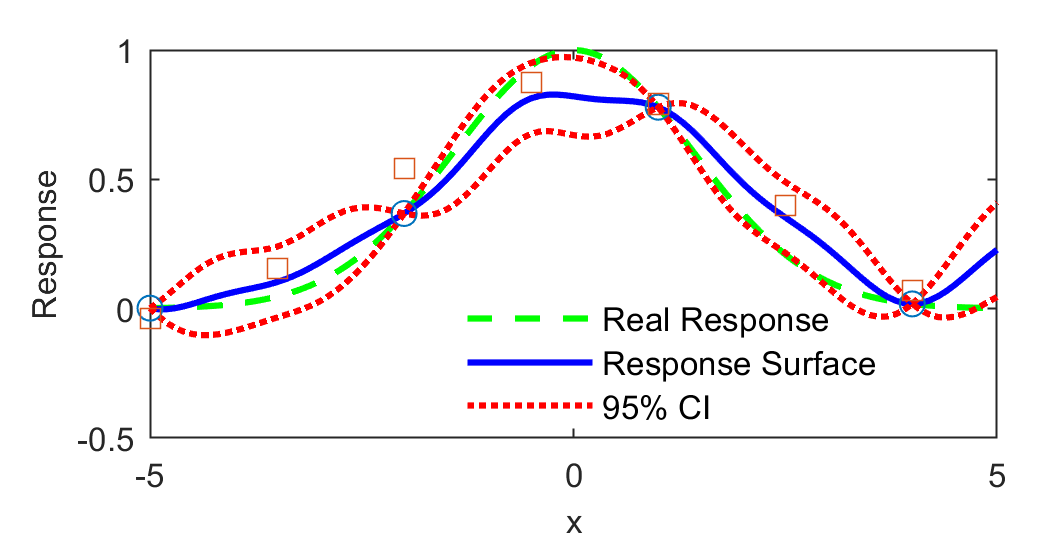}
      \caption{Response surface constructed from the experiments from the performance function $g(x)$ and the higher fidelity model $h_2(x)$. The blue circles represent the experiments from $g(x)$, the orange squares represent the experiments from $h_2(x)$.}
      \label{fig:exp1_2l}
\end{figure}

Now we start to consider the experiment results from the models $h_1,h_2$. We first consider the results from the higher fidelity model $h_2$ and use the proposed model to construct a response surface as described in Section \ref{sec:cokriging}. Fig. \ref{fig:exp1_2l} shows the response surface of the multi-fidelity model. The data points from $h_2$ are represented as orange squares. The resulting response surface is closer to the real function than the Kriging model (in Fig. \ref{fig:exp1_1l}), because the mean (blue solid line) has a similar shape and the confidence interval (red dot line) almost contains the real function everywhere in the region. This improvement is also confirmed by the mean squared error (MSE) of the mean of the response surfaces. The MSE decreased from 0.0572 (the Kriging model in Fig \ref{fig:exp1_1l}) to 0.0093 (the multi-fidelity model in Fig \ref{fig:exp1_2l}). Because the tail region $[4,5]$ does not have any experiments, the shape of the true function is still not well captured (the confidence interval still contains the real function in most part).

\begin{figure}[t]
      \centering
      \includegraphics[width=\linewidth]{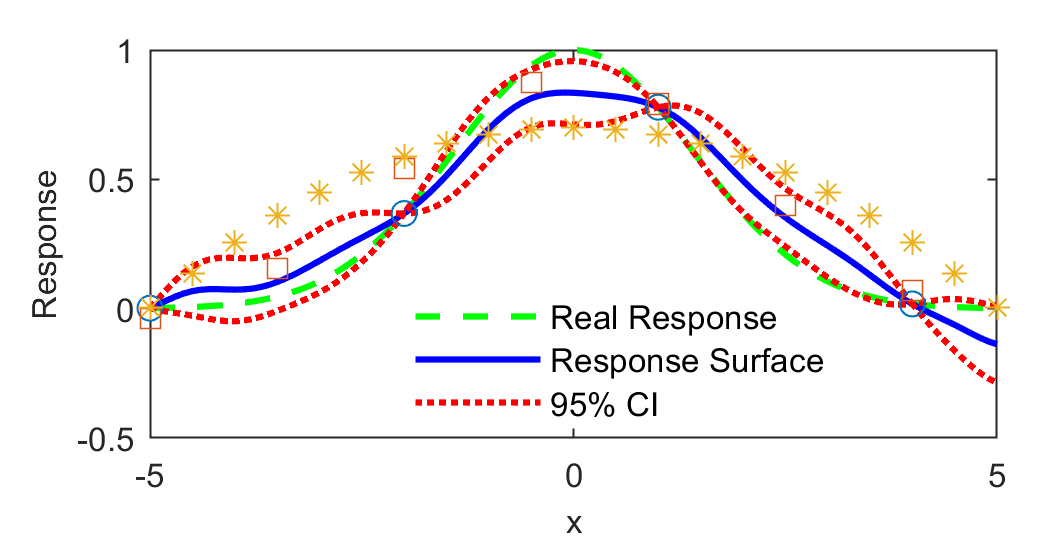}
      \caption{Response surface constructed from the experiments from the performance function $g(x)$, the higher fidelity model $h_2(x)$, and the lower fidelity model $h_1(x)$. The blue circles represent the experiments from $g(x)$, the orange squares represent the experiments from $h_2(x)$, the yellow asterisks represent the experiments from $h_1(x)$.}
      \label{fig:exp1_3l}
\end{figure}

We then further take the experiments from the lower fidelity model into consideration. We construct a multi-fidelity model with experiments from all models (from $g,h_1,h_2$). The response surface is shown in Fig. \ref{fig:exp1_3l}, where we use yellow asterisk to represent the experiments from $h_1$. Compared to the response surface in Fig \ref{fig:exp1_2l}, this response surface has better prediction in the region $[4,5]$, while it has a similar response in rest of the design space. The improvement of the tail region further decreased the MSE to 0.0087 (compare to 0.0093 in Fig \ref{fig:exp1_2l}).

The comparison of the three response surfaces shows us how the proposed model improves our prediction with lower fidelity models. In the region that experiments of higher fidelity models is not available, the model utilizes the information provided by the lower fidelity models. On the other hand, involving lower fidelity model experiments does not change the region with sufficient higher fidelity models.

\subsection{Implementation to AV Test Scenario}

Here we study the test scenario example we described in Section \ref{sec:setting}. In the test scenario, we want to study the performance of a test AV in the driving environment. The performance function of interest $g(x)$ is the minimum range of the test vehicle and the cut-in vehicle. Note that the input of the function $x$ is consisted of three variables $x=[v,\dot{R},R]$.

To implement the proposed model, we use experiment results from two models. We first select design points for input $x$ using a mesh grid design that has $v$ from 5 m/s to 35 m/s with a 2 m/s interval, $\dot{R}$ from 0 m/s to 30 m/s with a 2 m/s interval, $1/R$ from 0.1 $m^{-1}$ to 1 $m^{-1}$ with a 0.1 $m^{-1}$ interval. Let us denote the design point set as $X$ and $X$ contains 2,560 design points. We collect experiment results on these design points $X$ from the real performance function $g(x)$ and denote the experiment set as $D=(X,Y)$. We then randomly split the set into two, $D_1$ and $D_t$, where $D_1$ contains 1000 samples. We further extract 500 samples from $D_1$, and denote the obtained set as $D_2$. We then perturb the experiment results in $D_1$ by a uniform noise from $[-0.5,0.5]$. Now we consider $D_1$ as the data set from the lower fidelity model and $D_2$ as the data set from the higher fidelity model (or real function).

Similar to Section \ref{sec:toy}, we first use the higher fidelity data set $D_2$ to construct a Kriging model and then use $D_1$ and $D_2$ to construct a multi-fidelity model. We compare the two models to show the advantage of taking lower fidelity experiments into consideration. In this case, we use $D_t$ as the test data set to compute the MSE of the response surface mean of the two models. The result shows that the proposed model reduces the MSE to 2.3836 from 3.3948 of the Kriging model. 





 


\bibliographystyle{IEEEtran}
\bibliography{bibi.bib}


\end{document}